# Giant Interaction-Induced Gap and Electronic Phases in Rhombohedral Trilayer Graphene

Y. Lee, D. Tran, K. Myhro, J. Velasco Jr., N. Gillgren, C. N. Lau[†], Y. Barlas
Department of Physics and Astronomy, University of California, Riverside, Riverside, CA 91765
J.M. Poumirol, D. Smirnov
National High Magnetic Field Laboratory, Tallahassee, FL 32310
F. Guinea
ICMM - CSIC, E-28049 Madrid, Spain

**Due to their unique electron dispersion and lack of a Fermi surface, Coulomb interactions in undoped two-dimensional Dirac systems, such as single, bi- and tri-layer graphene, can be marginal or relevant. Relevant interactions can result in spontaneous symmetry breaking, which is responsible for a large class of physical phenomena ranging from mass generation in high energy physics to correlated states such as superconductivity and magnetism in condensed matter. Here, using transport measurements, we show that rhombohedral-stacked trilayer graphene (r-TLG) offers a simple, yet novel and tunable, platform for study of various phases with spontaneous or field-induced broken symmetries. Here, we show that, contrary to predictions by tight-binding calculations, rhombohedral-stacked trilayer graphene (r-TLG) is an intrinsic insulator, with a giant interaction-induced gap $\Delta \sim 42$ meV. This insulating state is a spontaneous layer antiferromagnetic with broken time reversal symmetry, and can be suppressed by increasing charge density $n$, an interlayer potential $U_\perp$, a parallel magnetic field, or a critical temperature $T_c \sim 38$K. This gapped collective state can be explored for switches with low input power and high on/off ratio.**

Few-layer graphene has emerged as unique systems for exploration of physical processes confined to two dimensions, as well as for band gap engieering. In the single particle picture, r-TLG hosts chiral charge carriers with Berry's phase $3\pi$, and an energy-momentum dispersions $\varepsilon(k) \sim k^3$, thus they are gapless semiconductors (Fig. 1a inset)[1-5]. Upon applying a potential difference $U_\perp$ between the outmost layers, the band structure adopts a tunable gap that scales with $U_\perp$ [1-7]. This single particle picture fails, however, close to the charge neutrality point (CNP), where the diverging density of states $D(\epsilon) \propto \varepsilon^{-1/3}$ and the rapidly increasing interaction strength $\alpha_s \propto n^{-1}$ lead to strong electronic interactions (here $n$ is the charge density). Thus the gapless semiconductor is expected to give way to phases with spontaneous broken symmetries, such as layer antiferromagnetic (LAF) and quantum anomalous Hall states with broken time reversal symmetry, and nematic states with broken rotation symmetry[8-13]. Such states are similar to those in bilayer graphene (BLG)[14-20]; however, many of these states, if present, are also expected to be more robust than BLG, due to the much larger $\alpha_s$.

The presence of gapped states and the rich phase diagram offered by r-TLG has not been explored to date, due to the lack of high quality dual-gated devices that allow independent control of both $n$ and $U_\perp$ [21]. For instance, though an insulating state with ~6 meV gap has been observed in singly-gated r-TLG near the CNP and attributed to interaction-induced effects[22], skepticism remains due to the absence of $U_\perp$-dependent studies.

---

[†] Email: jeanie.lau@ucr.edu

Here we present transport measurements on high mobility dual-gated r-TLG devices that allow independent modulation of both $n$ and $U_\perp$. At $n=U_\perp=0$, we observe a giant interaction-induced gap $\Delta\sim 42$ meV, which is almost an order of magnitude larger than that previously measured[22]. The gap is suppressed by $U_\perp$ of either polarity, by an in-plane magnetic field, or by a critical temperature $T_c\sim$ 34 K with temperature dependence $\Delta(T)\sim\sqrt{1-T/T_c}$ that is characteristic of phase transitions. Our results indicate that the gapped insulating state is a layer antiferromagnetic (LAF) state with broken time reversal symmetry. We expect that, upon application of large $U_\perp$, it transitions to quantum valley Hall state with broken inversion symmetry; when large and $B_\parallel$ is applied, the electronic state crosses over to the canted antiferromagnet (CAF) and eventually to ferromagnetic state.

r-TLG sheets on Si/SiO$_2$ are identified by Raman spectroscopy[23, 24] (Fig. 1a), Suspended dual-gated devices (Fig. 1b) with mobility as high as 90,000 cm$^2$/Vs are measured in He$^3$ refrigerators[25]. All measurements are taken at $T$=260mK, unless otherwise specified.

At low temperatures and $B$=0, r-TLG devices become insulating in the vicinity of the CNP. In Fig. 1c, the two-terminal differential conductance $G=dI/dV$ from device 1 is plotted as $n$ and $U_\perp$. At $U_\perp=0$, as $n$ approaches the CNP, $G(n)$ decreases by more than 5 orders of magnitude to $<10^{-4}$ $e^2/h$, where $e$ is electron charge and $h$ is Planck's constant (Fig. 1d). This insulating state at the CNP is extremely robust, as it persists for the entire gate ranges, up to the highest applied $U_\perp$ (~65 mV).

Transport spectroscopy[17, 26] at $n=U_\perp=0$ reveals intriguing features: at small source-drain bias V, the device stays insulating; but as $V$ increases ±42 mV, $G$ rises by more than 6 orders of magnitude to extremely sharp peaks, then decreases to ~15 $e^2/h$ for larger $V$ (Fig. 2a). Such a $G(V)$ curve strongly resembles the density of state of a gapped phase, suggesting the presence of an energy gap $\Delta\sim$ 42 meV at $n=U_\perp=B=0$. With increasing charge density $n$, the gap diminishes and eventually disappears entirely (Fig. 2b). At $n=3\times 10^{11}$ cm$^{-2}$, $G(V)$ is flat, indicating that r-TLG becomes gapless at high density.

The gapped, insulating state near the CNP in the absence of external fields is unexpected from tight-binding calculations, but instead suggests a phase arising from electronic interactions with spontaneous broken symmetries. The magnitude of the gap, ~42 meV, is exceedingly large for an interaction-induced state. It is more than an order of magnitude larger than that found in BLG[17, 19], reflecting the divergent nature of the density of states and strong electronic interactions in r-TLG at the CNP. It also a factor of 7 larger than that previously observed in singly-gated devices[22], likely due to improved device geometry and quality.

To further establish the magnitude of the gap, we examine temperature dependence of $G(V)$ at $n=U_\perp=0$ (Fig. 2c). Fig. 2d plots minimum conductance $G_{min}=G(V=0)$ as a function of $T$. At high temperatures $T>40$K, the device is conductive – $G_{min}$ ~15 $e^2/h$ with a small linear $T$-dependence. The $G(V)$ curves are approximately constant, similar to that of a conventional resistor. However, when $T<\sim 40$K, $G_{min}$ drops precipitously and become insulating for $T<30$K, and $G(V)$ curves develop prominent peaks at finite $V$. In the transition region $30<T<40$K, the $G_{min}(T)$ curve is well-described by the thermal activation equation, $G_{min}=G_0 e^{-\Delta/2k_B T}$ (Fig. 2e), where $k_B$ is the Boltzmann's constant and $\Delta\sim$43 meV is obtained as a fitting parameter. This is in excellent agreement with the value of $\Delta$ obtained from $G(V)$ curves at $T$=300 mK, thus confirming the presence of an insulating state with ~42±1 meV gap.

Using $G(V)$ curves, we can also directly measure the evolution of $\Delta$ (taken as half of the peak-to-peak separation in $V$) as a function of $T$. As shown in Fig. 2f, $\Delta$ is almost constant for

$T$<10 K, but drops precipitously for $T$>30K. This behavior is characteristic of order parameters during phase transitions in mean field theories. Thus we fit $\Delta(T)$ to the function[27]

$$\Delta(T) = \Delta(0)\left[A\left(1-\frac{T}{T_c}\right)+B\left(1-\frac{T}{T_c}\right)^2\right]^{1/2} \qquad (1)$$

where $T_c$ is the critical temperature. Eq. (1) reduces to the usual mean –field functional form $\sqrt{1-\frac{T}{T_c}}$ for $T/T_c$ sufficiently close to 1, and the second term $\left(1-\frac{T}{T_c}\right)^2$ is inserted to capture the vanishingly small dependence on $T$ as $T\to 0$. Excellent agreement with data is obtained, yielding $A$=2.0, $B$=-1.0 and $T_c$=34K. The energy scale of the gap, $\Delta(0)/k_B$=500K, which is much larger than that associated with the critical temperature, signifies that this insulating state observed at the CNP is a correlated phase.

To elucidate the nature of this correlated phase, we examine its dependence on external electric and magnetic fields. Fig. 3a displays $G$ as a function of $V$ and $U_\perp$ at $n$=0. As $U_\perp$ is the externally imposed potential bias, it will be heavily screened due to r-TLG's large density of states near the CNP[28-30]. Thus we expect the screened interlayer potential bias $U_\perp^s \ll U_\perp$. Using a simplified two-band Hamiltonian for r-TLG, we self-consistently calculate $U_\perp^s$ for given values of $n$ and $U_\perp$ [25], assuming that the dielectric constant of r-TLG is 1. The screening-corrected data $G(V, U_\perp^s)$ are shown in Fig. 2b. The sharp peaks in $G(V)$, i.e. the gap edges, appear as red curves that separate the insulating (dark blue) and conductive (light blue) regions in Fig. 2a and 2b. $\Delta$ decreases symmetrically and linearly with applied $U_\perp^s$ of either polarity, to ~ 30 meV at $|U_\perp|$=50meV or $|U_\perp^s|$=3mV (Fig. 3b), though not yet completely closed at the largest applied $|U_\perp|$. (In other devices with lower mobility, we also observe the complete suppression of the insulating state and an increase in $G$ with increasing $|U_\perp|$, as shown in Extended Data Fig. 3).

Finally, we investigate the response of the gapped correlated state to in-plane magnetic field $B_{||}$. Fig. 3d plots the $G(V,B_{||})$ at $n=U_\perp$=0. Interestingly, the gap is partially suppressed by large $B_{||}$: at 31T, $\Delta$ is reduced by ~ 5meV, suggesting that the insulating state is spin-polarized.

To summarize our experimental findings: we observe an insulating state in r-TLG at $n=U_\perp=B$=0, with an energy gap $\Delta(T=0)$~ 42 meV. This gap can be suppressed by increasing charge density $n$, a critical temperature $T_c$~34K, by an interlayer potential $U_\perp$ of either polarity, and by an in-plane magnetic field. Among the many possible correlated phases in r-TLG[8-13], only LAF, in which the top and bottom layers have equal number of electrons with opposite spin polarization, is consistent with our experimental observations. For instance, the presence of an energy gap eliminates the mirror-breaking, inversion breaking, interlayer current density wave or layer polarization density wave states [12], and the zero conductance eliminates the superconductor, quantum spin Hall and quantum anomalous Hall states that host finite (or even infinite) conductance. Furthermore, the symmetrical suppression of the gap by $U_\perp$ of either polarity suggests that charges in the insulating state are layer-balanced, since the device would otherwise exhibit opposite dependence on $U_\perp$ of opposite polarities. This excludes all layer-polarized states, including the quantum valley Hall and layer polarization density wave states, and any single particle state that arises from inadvertent doping of one of the surface layers.

Thus we identify LAF with broken time reversal and spin rotation symmetries as the intrinsic ground state in r-TLG. A simple mean field estimate yields[25]

$$\Delta \approx 2\times\left(\frac{c}{\pi\sqrt{3}}\right)^3 \frac{\gamma_1^4 U^3}{\gamma_0^3} \qquad (2)$$

where $\gamma_0 \approx 2.7$ eV, $\gamma_1 \approx 0.4$ eV are tight binding parameters, $c \approx 2.8$, and $U$ is the Hubbard onsite interaction. Using these parameters, and substituting the experimentally obtained value $\Delta$=42meV, we obtain $U$~13 eV, not too different from theoretically predicated values of 5-10 eV[31-33]. Alternatively, the gap can be further enhanced by exchange processes associated to the long range part of the interaction[31].

The proposed phase diagram for charge neutral r-TLG, together with schematics for electron configurations, is summarized in Fig. 4. In the absence of external fields, a charge neutral r-TLG is an LAF with broken time reversal and spin rotation symmetries. Increasing $U_\perp$ of either polarity pushes electrons to one of the surface layers and suppresses the gap. For sufficiently large $|U_\perp|$, all charges reside in either the top or bottom layer, giving rise to a quantum valley Hall (QVH) insulator with broken inversion symmetry. On the other hand, as $B_{||}$ increases from 0, the competition between the Zeeman and the exchange energies tilts the electron spins, and r-TLG crosses over to the CAF phase. For very large $B_{||}$, we expect that the electrons eventually form a ferromagnet (F) with counter-propagating edge states and conductance ~ $6e^2/h$.

In conclusion, we have demonstrated that the unusually large density of states and competing symmetries in r-TLG give rise to an insulating layer antiferromagnetic phase in the absence of external fields. This LAF state appears to be the thermodynamic ground state, and has a large gap ~42 meV. Transitions to canted anti-ferromagnetic, ferromagnetic and layer polarized states can be tuned by $U_\perp$, $B_{||}$, or $n$. Finally, our results shed light on other layered 2D systems such as few-layer graphene and double-layer QH systems, and with implications for graphene-based electronics and optoelectronics.

**References and Notes**

1. M. Koshino and E. McCann, Phys. Rev. B **80**, 165409 (2009).
2. F. Guinea, A. H. Castro Neto, and N. M. R. Peres, Phys. Rev. B **73**, 245426 (2006).
3. A. A. Avetisyan, B. Partoens, and F. M. Peeters, Phys. Rev. B **81**, 115432 (2010).
4. M. Aoki and H. Amawashi, Sol. State Commun. **142**, 123 (2007).
5. F. Zhang, B. Sahu, H. K. Min, and A. H. MacDonald, Phys. Rev. B **82**, 035409 (2010).
6. K. Zou, F. Zhang, C. Capp, A. H. MacDonald, and J. Zhu, Nano Lett. **13**, 369 (2013).
7. C. H. Lui, Z. Li, K. F. Mak, E. Cappelluti, and T. F. Heinz, Nat. Phys. **7**, 944 (2011).
8. E. V. Castro, M. Pilar Lopez-Sancho, and M. A. H. Vozmediano, Solid State Commun. **152**, 1483 (2012).
9. R. Olsen, R. van Gelderen, and C. M. Smith, Phys. Rev. B **87** (2013).
10. H. Liu, H. Jiang, X. C. Xie, and Q.-f. Sun, Phys. Rev. B **86**, 085441 (2012).
11. J. Jung and A. H. MacDonald, preprint, arXiv:1208.0116 (2012).
12. V. Cvetkovic and O. Vafek, preprint, arXiv:1210.4923 (2012).
13. M. M. Scherer, S. Uebelacker, D. D. Scherer, and C. Honerkamp, Phys. Rev. B **86**, 155415 (2012).
14. J. Martin, B. E. Feldman, R. T. Weitz, M. T. Allen, and A. Yacoby, Phys. Rev. Lett. **105**, 256806 (2010).
15. R. T. Weitz, M. T. Allen, B. E. Feldman, J. Martin, and A. Yacoby, Science **330**, 812 (2010).

16. W. Bao, J. Velasco, Jr., F. Zhang, L. Jing, B. Standley, D. Smirnov, M. Bockrath, A. H. MacDonald, and C. N. Lau, Proc. Nat. Acad. Sci. **109**, 10802 (2012).
17. J. Velasco, et al., Nature Nanotechnol. **7**, 156 (2012).
18. A. Veligura, H. J. van Elferen, N. Tombros, J. C. Maan, U. Zeitler, and B. J. van Wees, Phys. Rev. B **85**, 155412 (2012).
19. F. Freitag, J. Trbovic, M. Weiss, and C. Schonenberger, Phys. Rev. Lett. **108**, 076602 (2012).
20. A. S. Mayorov, et al., Science **333**, 860 (2011).
21. T. Khodkov, F. Withers, D. C. Hudson, M. F. Craciun, and S. Russo, Appl. Phys. Lett. **100**, 013114 (2012).
22. W. Bao, et al., Nat. Phys. **7**, 948 (2011).
23. C. H. Lui, Z. Li, Z. Chen, P. V. Klimov, L. E. Brus, and T. F. Heinz, Nano Lett. **11**, 164 (2011).
24. A. C. Ferrari, et al., Phys. Rev. Lett. **97**, 187401 (2006).
25. Supplmentary Information.
26. J. Velasco Jr., Y. Lee, Z. Zhao, L. Jing, P. Kratz, M. Bockrath, and C. N. Lau, Nano Lett. **14**, in press (2014).
27. R. A. Ferrell, Zeitschrift fur Physik **182**, 1 (1964).
28. R. van Gelderen, R. Olsen, and C. Morais Smith, preprint, arXiv:1304.5501 (2013).
29. M. Koshino, Physical Review B **81**, 125304 (2010).
30. H. Min, E. H. Hwang, and S. Das Sarma, Phys. Rev. B **86**, 081402 (2012).
31. T. O. Wehling, E. Şaşıoğlu, C. Friedrich, A. I. Lichtenstein, M. I. Katsnelson, and S. Blügel, Phys. Rev. Lett. **106**, 236805 (2011).
32. J. A. Verges, E. SanFabián, G. Chiappe, and E. Louis, **81**, 085120 (2010).
33. M. Schüler, M. Rösner, T. O. Wehling, A. I. Lichtenstein, and M. I. Katsnelson, Phys. Rev. Lett. **111**, 036601 (2013).

Acknowledgements: We thank R. Cote for discussions. This work is supported in part by DMEA H94003-10-2-1003, NSF DMR/1106358 and the FAME center that is one of the six STARNET centers. YL acknowledges the support by DOE. Part of this work was performed at NHMFL that is supported by NSF/DMR-0654118, the State of Florida, and DOE.

**Fig. 1. a.** Raman spectroscopy of a r-TLG sheet. **Inset:** Energy-momentum dispersion of r-TLG. **b.** SEM image of a dual-gated suspended TLG device. **c.** $G(U_\perp, n)$ in units of $e^2/h$. **d.** $G(n)$ at $U_\perp=0$. Note the logarithmic scale of $G$.

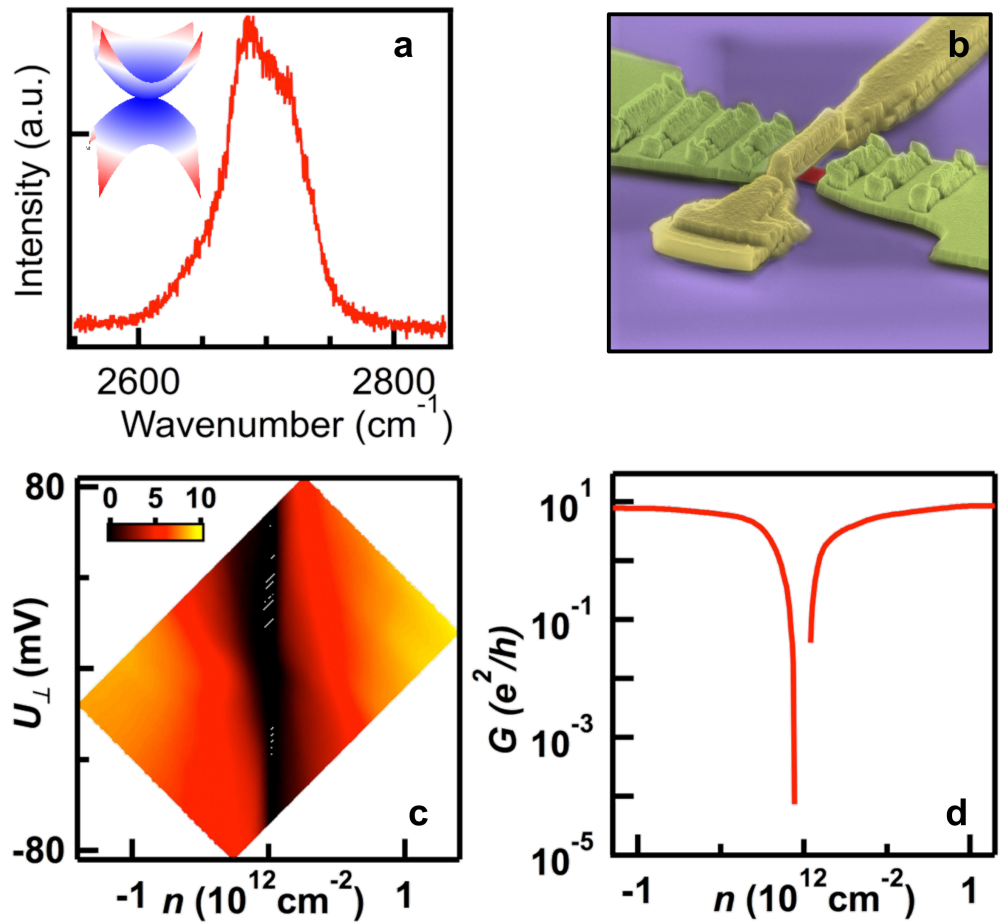

**Fig. 2.** Transport data at $B_\perp=0$ (a-b: Device 1; c-f: Device 2). **a.** $G(V)$ at $U_\perp=n=0$. **b.** $G(V)$ at $U_\perp=0$ and different $n$. **c.** $G_{min}(V)$ at $U_\perp=n=0$ and different temperatures. **d.** $G_{min}$ at $V=0$ vs. $T$. **ce** $G_{min}$ vs. $1/T$ in Arrhenius scale. The blue line is a fit to the equation $G_{min} = G_0 e^{-\Delta/2k_BT}$ for 30K<$T$<40K. **f.** Measured $\Delta$ as a function of $T$. The solid line is a fit to Eq. (1).

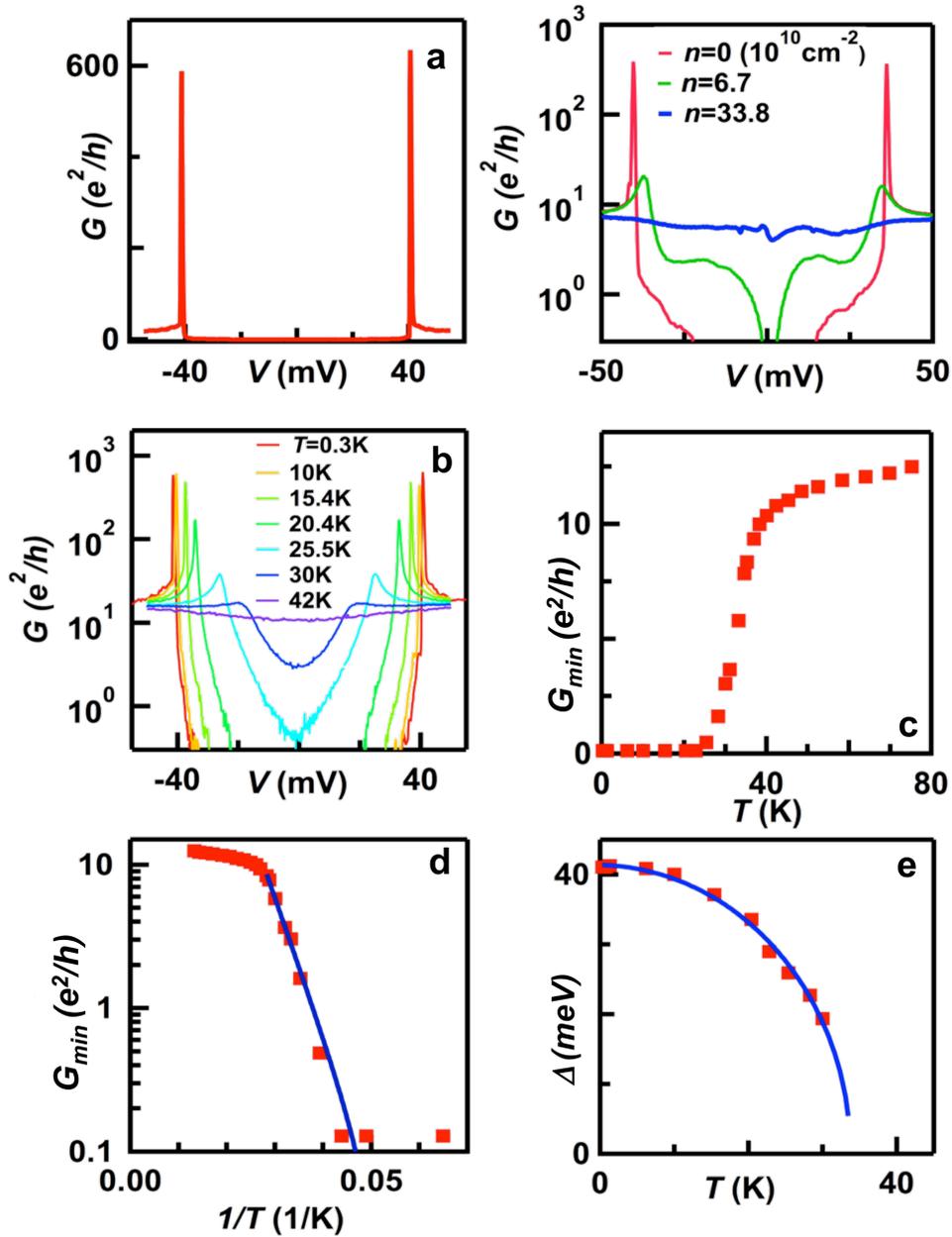

**Fig. 3.** Transport data at *n*=0 and finite $U_\perp$ and $B_\parallel$. **a-b.** $G(V, U_\perp)$ and $G(V, U_\perp^s)$ in units of $e^2/h$ from Device 1. **c.** Line traces $G(V)$ at $U_\perp$=0 and $U_\perp$=-50 mV. **d.** $G(V, B_\parallel)$ in units of $e^2/h$ from Device 2.

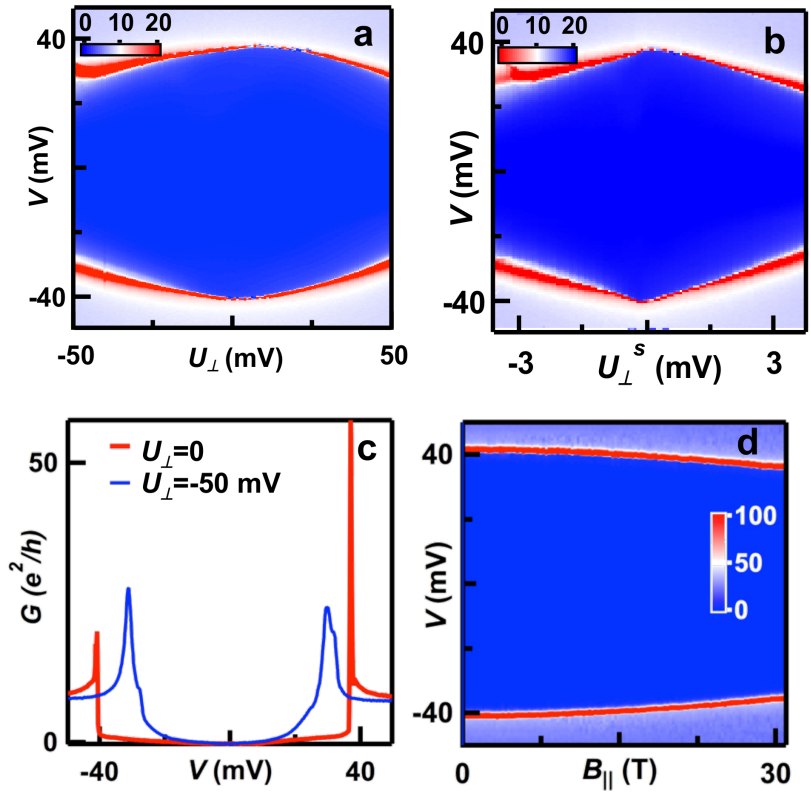

Fig. 4. Proposed phase diagram and schematics of electronic configurations for r-TLG. The blue and red arrows indicate charges from K and K' valleys, respectively.

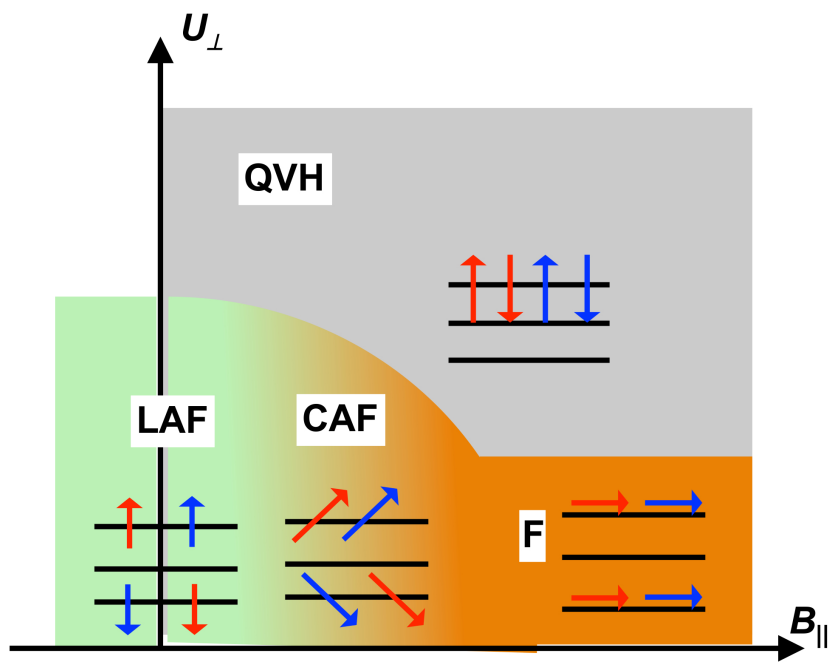